\documentclass[prl,showpacs,twocolumn,amsfonts,superscriptaddress]{revtex4}

\usepackage{graphicx}
\usepackage{bm}
\usepackage{amsmath,amssymb} 

\newcommand{\FuncD}{\mathcal{D}}
\newcommand{\HRS}{{\rm HRS}}

\begin{document}

\title{Random networks of cross-linked directed polymers}
\author{Stephan Ulrich} \affiliation{Institute for Theoretical
  Physics, Georg-August-Universit\"at G\"ottingen,
  Friedrich-Hund-Platz 1, 37077 G\"ottingen, Germany}

\author{Annette Zippelius} \affiliation{Institute for Theoretical
  Physics, Georg-August-Universit\"at G\"ottingen,
  Friedrich-Hund-Platz 1, 37077 G\"ottingen, Germany} \affiliation{Max
  Planck Institute for Dynamics \& Self-Organization, Bunsenstra{\ss}e
  10, 37073 G\"ottingen, Germany}

\author{Panayotis Benetatos} \affiliation{Institute for Theoretical
  Physics, Georg-August-Universit\"at G\"ottingen,
  Friedrich-Hund-Platz 1, 37077 G\"ottingen, Germany}  \affiliation{Theory of Condensed Matter
  Group, Cavendish Laboratory, University of Cambridge, 19 J.~J.~Thomson
  Avenue, Cambridge, CB3 0HE, United Kingdom}

\date{\today}  

\begin{abstract}
We explore the effect of random permanent cross-links on a system of directed polymers confined between two planes with their end points free to slide on them. We treat the cross-links as quenched disorder and we use a semimicroscopic replica field theory to study the structure and elasticity of this system. Upon increasing the cross-link density, we get a continuous gelation transition signaled by the emergence of a finite in-plane localization length. The distribution of localization length turns out to depend on the height along the preferred direction of the directed polymers. The gelation transition also gives rise to a finite in-plane shear modulus which we calculate and turns out to be universal, i.e., independent of the energy and length scales of the polymers and the cross-links. Using a symmetry argument, we show that cross-links of negligible extent along the preferred axis of the directed polymers do not cause any renormalization to the tilt modulus of the uncross-linked system.
\end{abstract}
\pacs{61.41.+e, 61.43.--j, 62.20.D--, 82.70.Gg}
\maketitle

\section{I. Introduction}
The statistical mechanics of directed polymers (DPs) has been a very active
field of research for more than twenty years
\cite{Kardar_book,Nelson_book}. The directed paths under study may
represent configurations of ``real'' extended one-dimensional objects
such as polymers \cite{Kamien}  and vortex lines in type-II
superconductors \cite{Blatter}, or may
represent configurations in abstract spaces such as those used to
model sequence alignment in bioinformatics \cite{Hwa}.  

Many physical systems consist of aligned extended one-dimensional
building blocks which can have crystalline or fluid-like order in the
transverse plane. Examples include columnar phases of DNA
\cite{Podgornik}, discotic \cite{Chandra} or micellar \cite{Safran}
liquid crystals, ferrofluids \cite{Rosen}, and electrorheological
fluids \cite{Halsey}. In addition, polymer brushes consisting of dense
flexible chains terminally anchored on a surface are characterized by
chain elongation in the direction of the surface normal
\cite{deGennes_brush}. Although the chains of polymer brushes can
assume backtracking conformations, under strong stretching they can be
viewed as directed strings of Pincus blobs \cite{Halperin_Zhulina}. In
recent years, there has been interest in cross-linked polymer brushes
because of promising technological applications \cite{Loveless,Xu}.

The effect of quenched disorder in the embedding medium on arrays of
interacting directed elastic lines has led to the prediction of a
whole zoo of glassy states in high-$T_c$ superconductors
\cite{Blatter}. In real polymer systems, irreversible cross-links can
be viewed as quenched disorder of a different type and their effect
can be studied using the tools of the statistical mechanics of
disordered systems \cite{DE}. A replica field theory has been used to
study the gelation transition due to permanent random cross-links in
systems comprised of Gaussian chains \cite{Castillo},
beads-and-springs \cite{SU_EPL}, dimers-and-springs \cite{Pfahl},
$p$-beine \cite{p-beine}, and wormlike chains \cite{PB_PRL}. A similar
field-theoretic approach to well-cross-linked macromolecular networks
has been developed by Panyukov and Rabin \cite{Pan_Rab1,Pan_Rab2} 

In this paper, we employ the same theoretical framework to study the
effect of permanent cross-links on a melt of flexible directed
polymers in a particularly simple geometry. The polymers are stretched
between two parallel flat surfaces with their ends free to slide on
them. We predict a gelation transition upon cross-linking associated
with the emergence of a finite localization length in the transverse
plane which depends on the distance from the boundary surfaces of the
slab.

Furthermore, we investigate mechanical properties of the system. Due
to the asymmetry of the system, one has to distinguish between
\emph{tilt modulus} and \emph{shear modulus}; the first one describes
the resistance to shear of the boundaries in the preferred chain
direction and the latter of the perpendicular boundaries. Remarkably,
the tilt modulus remains completely unaffected upon cross-linking with
cross-links of negligible extent in the aligning direction.

The paper is organized as follows. We present our model in Sec.~II. In Sec.~III, we define and calculate the tilt modulus for the cross-linking geometry of our model. We construct a replica field theory and obtain the gelation transition in Sec.~IV. The shear modulus is discussed in Sec.~V. We summarize in Sec.~VI.

\section{II. Model} 
\begin{figure}
\includegraphics[viewport=0 35 527 792,angle=270,width=0.46\textwidth]{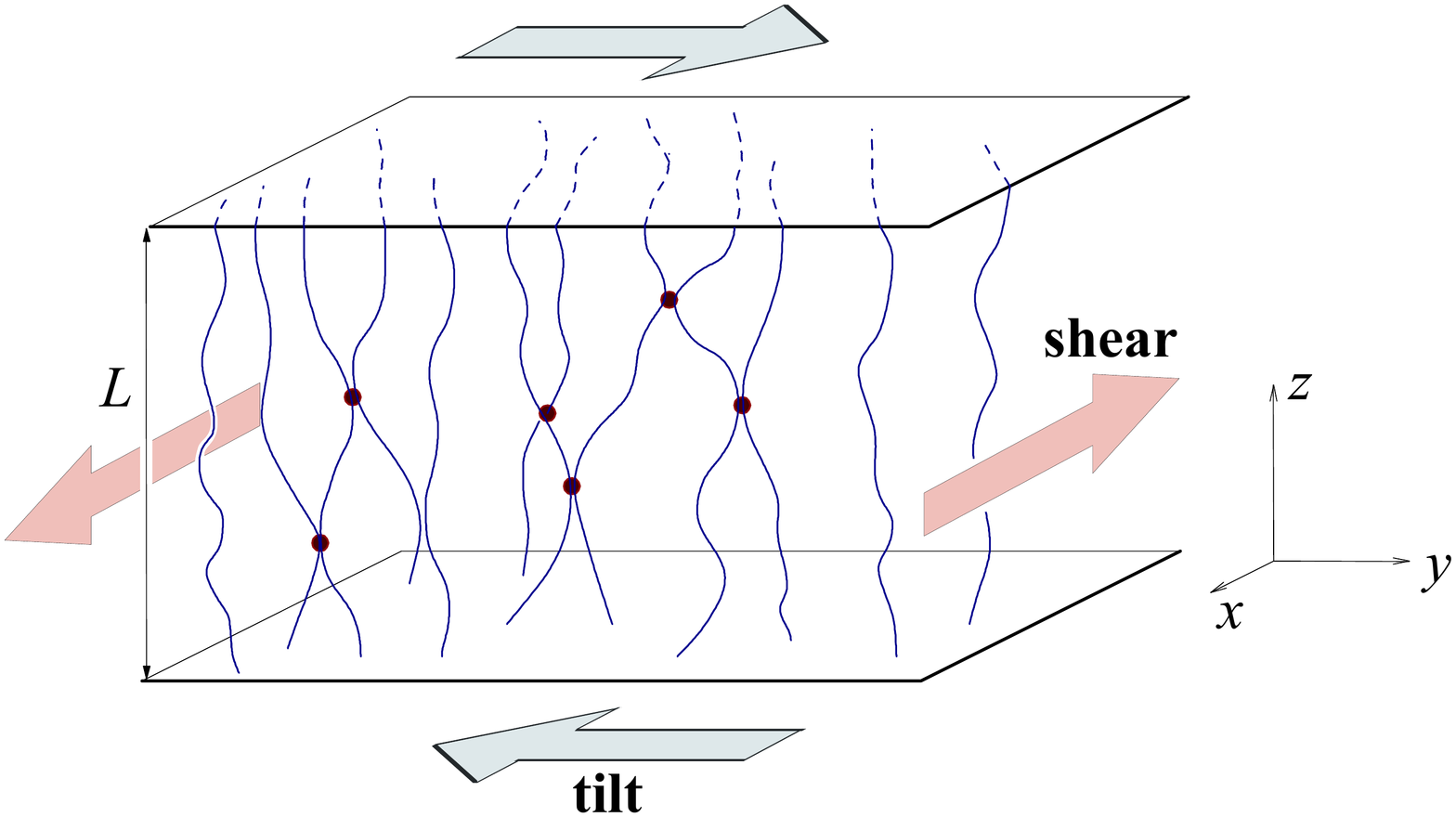}
\caption{(Color online) A schematic diagram of directed polymers in a slab of thickness $L$. $z$ is the preferred direction and we refer to $x,y$ as the \emph{transverse} or \emph{in-plane} direction.
\label{schematic}}
\end{figure}

We consider $N$ directed polymers stretched between two planes spaced
a distance $L$ apart. The end points of the polymers are free to slide
on the planes. Each polymer configuration is described by a curve
(path) ${\bf r}(z)=\bigl( x(z),y(z) \bigr)$, where $z \in [0,L]$ and
$z$ is the direction of alignment (Fig.~\ref{schematic}). By the
definition of directedness, these paths exclude loops and
overhangs. The areal density  of the system in the $xy$ plane is $N/A$. We assume free boundary conditions at $z=0$ and $z=L$,
allowing the polymer ends to assume any arbitrary position on the
corresponding planes with any slope. In the absence of cross-links,
the effective free-energy functional (``Hamiltonian'') of the directed
polymers consists of two terms,
\begin{align}
\label{Ham1}
{\cal H}_0 \{{\bf r}_i(z)\}
&=\sum_{i=1}^{N}\frac{\sigma}{2}\int_0^L dz \left(\frac{d{\bf r}_i(z)}{dz}\right)^{\!2} \nonumber\\
&+ \;\sum_{i<j}\frac{\lambda}{2}\int_0^L dz \, \delta\big({\bf r}_i(z)-{\bf r}_j(z)\big), \hspace{15pt}
\end{align}
where the first term penalizes tilting away from the $z$ direction
with $\sigma$ being the effective line tension and the second
term is an excluded volume interaction.

A physical system where the effective free energy of Eq.~(\ref{Ham1})
can be realized is that of strongly stretched wormlike chains of contour length
$L$. If the two plates are held apart by a pressure $P$ and the areal
density (on the $xy$ plane) of the polymers is high, each polymer will
be stretched by a tension ${\cal F}\approx PA/N$. A strong tension allows only
weakly tilting configurations. In this case, the aligning part of the
free energy reads
\begin{align}
\label{Ham_al1}
{\cal H}_{al}^{(1)} \{{\bf r}_i(z)\}
&=\sum_{i=1}^{N}\frac{\cal F}{2}\int_0^L dz  \left(\frac{d{\bf r}_i(z)}{dz}\right)^2 \nonumber\\
&+\;\sum_{i=1}^{N}\frac{\kappa}{2}\int_0^L dz\left(\frac{d^2{\bf r}_i(z)}{dz^2}\right)^2,
\end{align}
where ${\kappa}$ is the bending stiffness of the wormlike chains
related to their persistence length $L_p$ via ${\kappa}=L_p k_{\rm B} T$. For
${\cal F}\gg (k_{\rm B} T)^2/{\kappa}$ and $L\gg L_p$, one can 
show \cite{deGennes} that the bending term on the rhs of the previous
equation can be neglected and the projection of the polymer on the
$xy$ plane behaves as a Gaussian chain. 

In Ref.~\cite{Halperin_Zhulina}, a realization of a stretched brush is
envisioned as \emph{ABA} triblock lamellae with selective cross-linking of
the \emph{A} blocks. In an analogous realization of our model, the \emph{A} blocks
would form fluid membranes.

Another physical realization of Eq.~(\ref{Ham1}) is that of wormlike
chains interacting with a strong nematic field
\cite{Nelson_book,Kamien}. If the chains are embedded in a nematic
solvent with very large Frank constants and the effective tension due
to the polymer-nematic interaction ${\sigma}$ is such that
${\sigma}\gg (k_{\rm B} T)^2/{\kappa}$, hairpins are negligible. If, in
addition, $L\gg L_p$, the bending stiffness can be neglected altogether
and the chains behave as directed flexible polymers.

In the system described by Eq.~(\ref{Ham1}), we introduce $M$ permanent
cross-links which restrict linked polymer segments
to remain within a distance of order $a$. Their effect is described by
an effective interaction
\begin{align}
  \label{Ham_xlink}
\frac{{\cal H}(\mathcal{C}_M)}{k_{\rm B} T} = \frac{1}{2a^2}\sum_{e=1}^M\big({\bf r}_{i_e}(z_e)
-{\bf r}_{j_e}(z_e)\big)^2.
\end{align}
$\mathcal{C}_M := \{i_e,j_e;z_e\}$ is a quenched configuration of $M$ cross-links identified by the
polymers $i_e,j_e$ involved and the cross-linking height $z_e$. For the sake of simplicity, we assume that the
cross-linking interaction is ``local'' in the $z$ direction and only
depends on the in-plane distance of the polymer segments. As we shall
show in Sec.~III, this assumption has profound consequences for the
elasticity of the cross-linked system. 

The partition function of the system for a specific realization of
cross-links, $\mathcal{C}_M$, reads as
\begin{eqnarray}
\label{Z_C_M}
Z(\mathcal{C}_M)=\left\langle\exp\!\left(-\frac{{\cal H}(\mathcal{C}_M)}{k_{\rm B} T}\right)\right\rangle,
\end{eqnarray}
where $\langle...\rangle$ denotes average over all polymer
configurations with Boltzmann weight $\exp(-{\cal H}_0/k_{\rm B} T)$. Physical
observables of interest can be calculated from the quenched-disorder
averaged free energy, $F=-k_{\rm B} T [\ln Z]$, where $[...]$ denotes
average over all realizations of random cross-links. We assume that the
number of cross-links can vary and a realization with $M$
cross-links follows the Deam-Edwards distribution \cite{DE}:
\begin{eqnarray}
\label{DE}
P(\mathcal{C}_M)\propto\frac{1}{M!}\left(\frac{{\mu}^2 A}{2N(2{\pi}a^2)}\right)^{\!M} Z(\mathcal{C}_M).
\end{eqnarray}
The parameter ${\mu}^2 = 2 {[M]}/{N}$ controls the average
number of cross-links per polymer, and the physical meaning of this
distribution is that polymer segments close to each other in the
un-cross-linked phase have a high probability of getting linked.

\section{III. Tilt Modulus}
On large length scales, an array of directed polymers can be described
as an elastic continuum with three elastic moduli: a shear, a bulk,
and a tilt modulus. The first two characterize deformations in the transverse
plane whereas the third characterizes the response to tilting away 
from the preferred axis. The elastic free energy of such a system was
proposed in the context of vortex-line arrays in type-II superconductors by
de Gennes and Matricon \cite{deG-Mat}:
\begin{align}
\label{F_el}
F_{el}=\frac{1}{2}\int\frac{d^2 q}{(2{\pi})^2} &\int\frac{d
  q_z}{2{\pi}}\Big\{ (Kq_z^2+Gq^2) \, |{\bf u}({\bf q},q_z)|^2\nonumber\\
&+ Bq_\mu q_\nu u_\mu({\bf q},q_z) u_\nu(-{\bf q},-q_z)\Big\},
\end{align}
where $K$, $G$, and $B$ are, respectively, the tilt, shear, and bulk
modulus. $\mu,\nu$ are Cartesian indices in the $xy$ plane. ${\bf
  u}({\bf q},q_z)$ is the Fourier transform of the two-component
displacement field which parametrizes the elastic distortion of
the vortex-line array.

In order to measure the tilt modulus, we consider a small force, ${\bf f}$,
applied at the upper end point of each polymer and the opposite force,
$-{\bf f}$, applied at the lower end point. The induced deformation is
measured by the average tilt field. The tilt field is defined as 
\begin{eqnarray}
\label{tilt_field}
{\bf t}({\bf r},z)=\sum_{i=1}^N\frac{d {\bf r}_i}{dz}\delta\big({\bf r}-{\bf r}_i(z)\big).
\end{eqnarray}
The energy of the system subject to a cross-link configuration $\mathcal{C}_M$
and to a tilting force ${\bf f}$ reads as
\begin{eqnarray}
\label{Ham_f}
{\cal H}_{\bf f}={\cal H}_0+{\cal H}(\mathcal{C}_M)-\int_0^L dz \int d^2r\; {\bf
  f}{\cdot} {\bf t}({\bf r},z).
\end{eqnarray}
To leading order in ${\bf f}$, the average tilt field is
\begin{eqnarray}
\label{resp1}
\frac{1}{LA}\int_0^L dz \int d^2 r \langle {\bf t}({\bf r},z)
\rangle_{\bf f}=\frac{1}{K}{\bf f},
\end{eqnarray}
where $\langle ... \rangle_{\bf f}$ denotes thermal average with a Boltzmann
weight corresponding to the energy functional ${\cal H}_{\bf f}$.
The tilt modulus, $K$, can be extracted from the partition function as
a static linear-response coefficient,
\begin{eqnarray}
\label{tilt_mod}
\frac{1}{K}\delta_{\mu\nu}=\frac{k_{\rm B} T A}{N^2 L} \frac{{\delta}^2}{\delta f_\mu \delta f_\nu}\ln Z_{\bf f}{\bigg\arrowvert}_{{\bf f}={\bf 0}}, 
\end{eqnarray}
where
\begin{eqnarray}
\label{Z_f}
Z_{\bf f} = \int {\cal D}\{{\bf r}_i(z)\} \exp\!\left(-\frac{{\cal H}_{\bf f}\{{\bf r}_i(z)\}}{k_{\rm B}T}\right).
\end{eqnarray}
In the path integral of the previous equation, we apply a ``Galilean''
transformation (where the height $z$ is viewed as a time-like
parameter) \cite{TN},
\begin{eqnarray}
\label{Galilean}
{\bf r}_i &\longrightarrow& {\bf r}'_i={\bf r}_i\;+\;\frac{{\bf f}}{\sigma}z\nonumber\\
z  &\longrightarrow& z'=z
\end{eqnarray}
which brings it to the form
\begin{eqnarray}
\label{Z_f1}
Z_{\bf f}=Z_{{\bf f}={\bf 0}} \exp\!\left(-\frac{NL}{2k_{\rm B}T \sigma} {\bf f}^2\right),
\end{eqnarray}
where $Z_{{\bf f}={\bf 0}}$ is the partition function without the external field ${\bf f}$. Equations (\ref{Z_f1})
and (\ref{tilt_mod}) yield
\begin{eqnarray}
\label{tilt_mod1}
K=\frac{N}{A}{\sigma}.
\end{eqnarray}
This result implies that the tilt modulus of a directed polymer array
with a specific realization of cross-links of the type described by
Eq.~(\ref{Ham_xlink}) is completely unaffected by
the cross-links and simply reduces to the single-polymer tension. Since
any realization of the quenched disorder associated with the
cross-links of this type would give the same result, we are spared the burden of
having to use replicas for the calculation of the average over disorder.   

The reason behind the particularly simple result for the tilt
modulus is the ``Galilean'' invariance of the interactions between
the polymers as well as of the boundary conditions. In the specific
model, both the excluded volume interaction and the cross-link 
interaction involve polymer segments at the same height $z$ and
therefore remain unchanged under a ``Galilean''
transformation. Real cross-linking molecules have a finite extent and
may link to polymer segments at different heights thus breaking the
``Galilean'' invariance. That would lead to a non-trivial renormalization of the
tilt modulus. In the limiting case of cross-links with negligible extent, our
model is a good approximation, and we expect the tilt modulus to remain
unchanged  and be given by
Eq.~(\ref{tilt_mod1}).

\section{IV. Gelation Transition}
The system of cross-linked directed polymers undergoes a gelation
transition as the number of cross-links per chain increases. Whereas in
the sol phase the DPs are free to move in the $xy$ plane like particles
in a two-dimensional fluid, the polymers' motion in the gel phase is restricted to
finite excursions around preferred positions. Thus there is a
localization transition in the $xy$ plane, similar to the gelation
transition in systems comprised of other building blocks in $d=3$ \cite{Castillo}. Since
the latter has been discussed extensively, we keep our discussion
short.


What is the \emph{order parameter} for the localization transition in the $xy$
plane? A point $z$ on  curve $i$, i.e., monomer $z$ on polymer $i$ in
a discretized model, is localized, if it has a nontrivial expectation
value 
\begin{equation}
\label{local_density}
\langle\delta({\bf x}-{\bf r}_i(z))\rangle\neq 1/V.
\end{equation}
If the particles are localized at random positions, as we expect for
the gel phase, then the density averaged over all particles
vanishes at any nonzero wave vector. A possible order parameter is the second moment of the local
density:
\begin{equation}
\Omega^{(2)}({\bf q},z)=\frac{1}{N}\sum_{i=1}^N
\bigr[\langle e^{i{\bf q}\cdot{\bf r}_i(z)}
\rangle \langle e^{-i{\bf q}\cdot{\bf r}_i(z)} \rangle\bigr]\;.
\end{equation}
In general, one polymer is cross-linked with a finite number of other
polymers and in fact close to the transition this number is
small. Hence there is no reason to expect that the local density
should obey Gaussian statistics, therefore we need all moments of the
local density to characterize the gel. This is achieved in the replica
formalism by introducing $n$ copies, one for each thermal
expectation value. The order parameter in the replica theory
\begin{equation}
\label{order_parameter}
  \Omega({\bf x}_1...{\bf x}_n,z)=\frac{1}{N}\sum_{i=1}^N
\bigl[\langle \delta({\bf x}_1-{\bf r}_i(z))\rangle \cdots
\langle \delta({\bf x}_n-{\bf r}_i(z))\rangle\bigr], \nonumber
\end{equation}
captures all moments of the local density and hence characterizes the
structure completely. 

The average over the quenched realizations of cross-links, $\mathcal{C}_M$, is done with
help of the replica trick. 
The disorder averaged free energy $F=-k_{\rm B} T [\ln Z]=\lim_{n\to 0}
({\cal{Z}}_{n+1}-{\cal{Z}}_1)/(n{\cal{Z}}_1)$ is represented in terms of
$n$ noninteracting copies of the system together with one additional
replica to account for the distribution $P(\mathcal{C}_M)$ of
Eq.~(\ref{DE}) which is proportional to the partition function.
The replicated partition function is represented as a functional
integral over collective fields $\Omega(\hat{q},z)$,
\begin{subequations}\label{eq:ZnpeAll}
\begin{align}
{\cal{Z}}_{n+1} =& \int\FuncD \Omega \, e^{-{N\!f_{n+1}}}\;, \label{eq:Znpe}\\
\label{free_energy}
  f_{n+1}(\Omega)  =&\;  \phi^n \frac{\mu^2}{2L} \int_0^L \! 
  dz \!\sum_{\hat{q} \in \HRS} |\Omega(\hat{q},z)|^2 \Delta(\hat{q})\nonumber\\
  &+ \frac{1}{2L} \int_0^L \! 
  dz \!\sum_{\hat{q} \in {\rm 1RS}} 
  |\Omega(\hat{q},z)|^2 \tilde{\lambda}(\hat{q})\nonumber\\
&- \ln \mathfrak{z} \;, 
\end{align}
\end{subequations}
with the single-polymer partition function
\begin{align}
  \mathfrak{z} = & \int  \FuncD \hat{r}(z) \, e^{-H_{0}^{(n+1)}} \label{eq:free_energy_z}\\
  &\times\exp\Biggl(\frac{ \phi^n\mu^2}{L} \int_0^L dz  \sum_{\hat{q}
    \in \HRS} \Delta(\hat{q})  \Omega(\hat{q},z) e^{-i\hat{q} \cdot \hat{r}(z)}\nonumber\\  
   &\hspace{34pt}+ \frac{i}{L} \int_0^L \! 
  dz \!\sum_{\hat{q} \in {\rm 1RS}} \tilde{\lambda}(\hat{q}) \Omega(\hat{q} ,z) e^{-i\hat{q} \cdot \hat{r}(z)} \Biggr) \;, \nonumber 
\end{align}
where 
\begin{align}
 \tilde{\lambda}(\hat{q}):= \lambda \frac{L N}{2A} - \phi^n {{\mu}^2\Delta(\hat{q})}\;.
\end{align}
To simplify the notation we have introduced hatted vectors, such as
$\hat{q}:=({\bf q}_0,{\bf q}_1,...{\bf q}_n)$ for $(n+1)$-fold replicated
vectors. We have also adopted units of energy such that $k_\text{B}T\equiv
1$. The harmonic potential
for the cross-links is reflected in
$\Delta(\hat{q})=\exp{(-a^2\hat{q}^2/2)}$ and $\phi=2\pi a^2/A$. The collective field
$\Omega$ is almost the order parameter, discussed above, except for
the zeroth replica which we have introduced to account for the
disorder average $[...]$ in Eq.~(\ref{order_parameter}).


Areal density fluctuations are  represented by $\Omega(\hat{q},z)$ with
$\hat{q}=({\bf 0},...{\bf q}_{\alpha},...{\bf 0})$, i.e., only one
nonzero component (1RS). These fluctuations are penalized by the excluded
volume interation. The stability of the liquid state in mean-field
approximation (uniform density) is controlled by the coefficient of
the quadratic term in the fluctuations. A sufficiently strong excluded
volume interaction such that $ \tilde{\lambda}(\hat{q}) \gg 1$
together with the positive definiteness of the kernel (in $z_1,z_2$)
$\langle e^{-i\hat{q} \cdot (\hat{r}(z_1)- \hat{r}(z_2))}
\rangle=\exp(-\hat{q}^2|z_1-z_2|/2\sigma)$ preclude a collapse of the
liquid state. Since the areal density fluctuations are noncritical, we only
consider the order parameter in the so-called higher replica sector
(HRS) consisting of vectors $\hat{q}$ with at least two nonzero
components.

The expectation value of the order-parameter field
\begin{equation}
\left\langle\Omega(\hat{x},z)\right\rangle_{f}=
\left\langle \delta (\hat{x}- \hat{r}(z)) \right\rangle_{f}
\end{equation}
has to be calculated self-consistently with the weight of
Eq.~(\ref{eq:ZnpeAll}). Here we restrict ourselves to the saddle-point
approximation $\delta f_{n+1}/\delta \Omega=0$.
As for the gel transition of random coils, the saddle-point equation
is solved exactly by the following ansatz for the order parameter: 
\begin{align}
\label{eq:dp:ansatzQ}
\Omega(\hat{q},z) =&\; (1-Q) \delta_{\hat{q},\hat{0}} \nonumber\\
&+ Q
\delta_{{\bf q}_{\parallel},{\bf 0}} 
\int_0^\infty \!\! d \xi^2  \mathcal{P}(\xi^2,z) 
\exp\!\left( - \frac{\hat{q}^2 \xi^2}{2}  \right).
\end{align}
Here $Q$ denotes the fraction of DPs in the infinite cluster and hence
$1-Q$ is the fraction of DPs in the fluid state, giving rise to the first
(trivial) contribution to the order parameter. On the other hand, the
localized particles are characterized by the localization length
$\xi$, which fluctuates not only from polymer to polymer but also
along one directed polymer giving rise to a distribution of localization length
\begin{equation}
\mathcal{P}(\xi^2,z)=\frac{1}{QN}\sum_{j\in Q}\bigl\langle\delta\bigl(\xi^2-\xi^2_j(z) \bigr) \bigr\rangle
\end{equation}
which depends on $z$. After averaging over the disorder the system
still has macroscopic translational invariance in the $xy$ plane. This
requires that ${\bf q}_{\parallel} := \sum_{\alpha=0} ^n{\bf q}_{\alpha} = 0$.

The solution of the saddle-point equation reveals a gel transition at a
critical cross-link concentration
$\mu^2=1$, when a macroscopic cluster of cross-linked DPs is formed.
The gel transition is signaled by a nonzero value of the gel fraction
$Q$. Close to the gel transition, $\epsilon=\mu^2-1\ll 1$ grows
continuously from zero:
$Q=2\epsilon+{\cal O}(\epsilon^2)$.
To discuss the distribution of localization length, we note that we
have several length scales in our system: the internal length of a
directed polymer, $L$, the length of a cross-link, $a$, and the radius
of gyration in the $xy$ plane, which is determined by
$l:=\sqrt{L/(2\sigma)}$. We expect that the latter will set the
scale for the localization length, introduce the abbreviation
$\theta=(2/3+a^2/l^2) l^2/(\xi^2 \epsilon)$ 
and consider the distribution of rescaled, inverse localization
length, $\pi(\theta,s=z/L)$, with 
\begin{equation}
\pi(\theta,s)d\theta=\mathcal{P}(\xi^2,Ls)d \xi^2.
\end{equation}
This function is the solution of
\begin{align} 
&( 1 +  2\epsilon) \pi(\theta,s) 
 =  ( 1 +  \epsilon ) \int_0^1 \! d s_1 \,  \pi(\theta,s_1)  \\
&\hspace{5pt} + \frac{\epsilon}{2/3+a^2/l^2} \int_0^1 \! d s_1 \, \partial_\theta 
\Bigl( \theta^{2}\pi(\theta,s_1) \Bigr)
 \bigl\{  2 |s - s_1| +  a^2/l^2   \bigr\} \nonumber \\ 
&\hspace{5pt} +  \epsilon  \int_0^1 \! d s_1 d s_2 \int_0^\theta  d\theta_1 \, 
\pi(\theta_1,s_1) \pi(\theta - \theta_1,s_2)
+ {\cal O}(\epsilon^2)  \;. \nonumber
\end{align}

\begin{figure}
\begin{minipage}[h]{.33\textwidth}
\includegraphics[width=.99\textwidth]{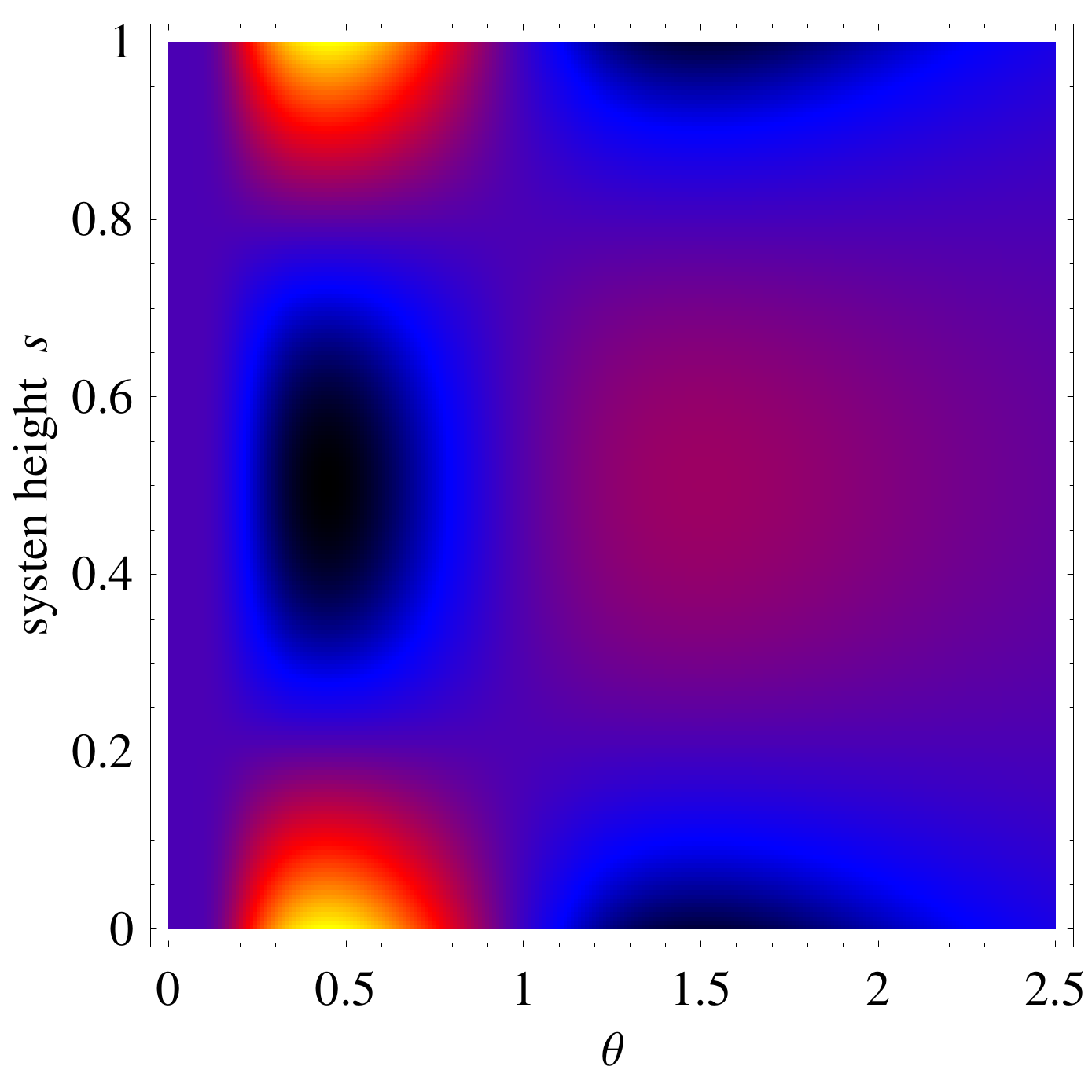}
\end{minipage}
\begin{minipage}[h]{.05\textwidth}
\includegraphics[height=5cm]{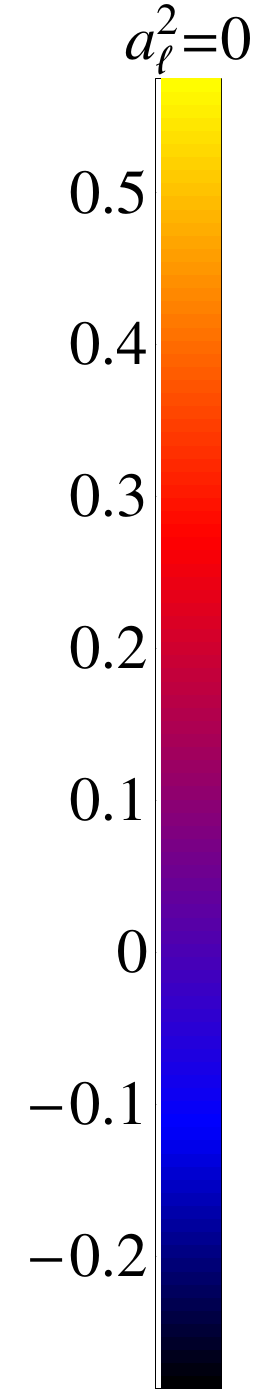} \\[5pt] \ 
\end{minipage}
\begin{minipage}[h]{.05\textwidth}
\includegraphics[height=5cm]{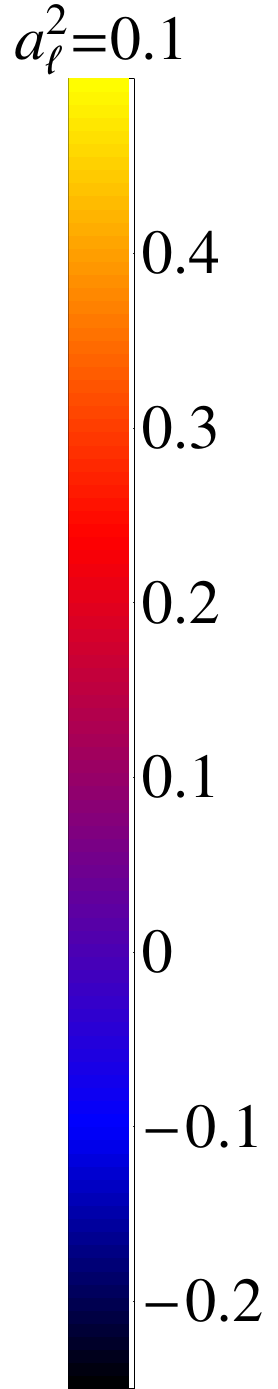} \\[5pt] \ 
\end{minipage}
\caption{(Color online) Height-dependent part, $\delta \pi(\theta,s)$, of the
  distribution of inverse localization length, $\theta$, as a function of $\theta
$ and $s:=z/L$. Due to the scaling behavior close to the sol-gel
transition, $\delta\pi(\theta,s)$ is normalized with $\epsilon$. $a_l^2$
is the ratio squared of the in-plane extent of a cross-link to the in-plane
radius of gyration of a free DP.}
\label{color_distrib}
\end{figure}

To gain a better understanding of the solution, we decompose the
distribution into its mean with respect to $s$, $\bar\pi(\theta)= \int
ds \,\pi(\theta,s)$ and a deviation: $\pi(\theta,s)=\bar\pi(\theta)+
\delta\pi(\theta,s) $. The mean, $\bar\pi(\theta)$, fulfills the same
equation as for isotropic gels \cite{Castillo}. The deviation is
small close to the gel point,
\begin{eqnarray}\label{eq:dp:DistEquationdp3} 
\delta\pi(\theta,s) 
& = &\epsilon w(s) \, \partial_\theta \Bigl( \theta^{2} \bar\pi(\theta) \Bigr)
  + {\cal O}(\epsilon^2)  \;, \\
 w(s)& =& \frac{ s^2 + (1-s)^2 - 2/3}{\qquad  a^2/l^2 + 2/3}  \;,
\end{eqnarray}
and furthermore controlled by the ratio of cross-link length to
in-plane radius of gyration: $a_l^2 := a^2/l^2$. The larger
the radius of gyration $l$, the more pronounced is the dependence on $s$.
We show the distribution for a
typical value $a_l^2=0.1$ in Fig.~\ref{color_distrib}. As one would
expect, localization is strongest in the middle of the directed polymer
and weaker at the boundaries. The variation across the length of the
DPs is stronger for larger localization length (small $\theta$).

\begin{figure}
\includegraphics[width=.33\textwidth]{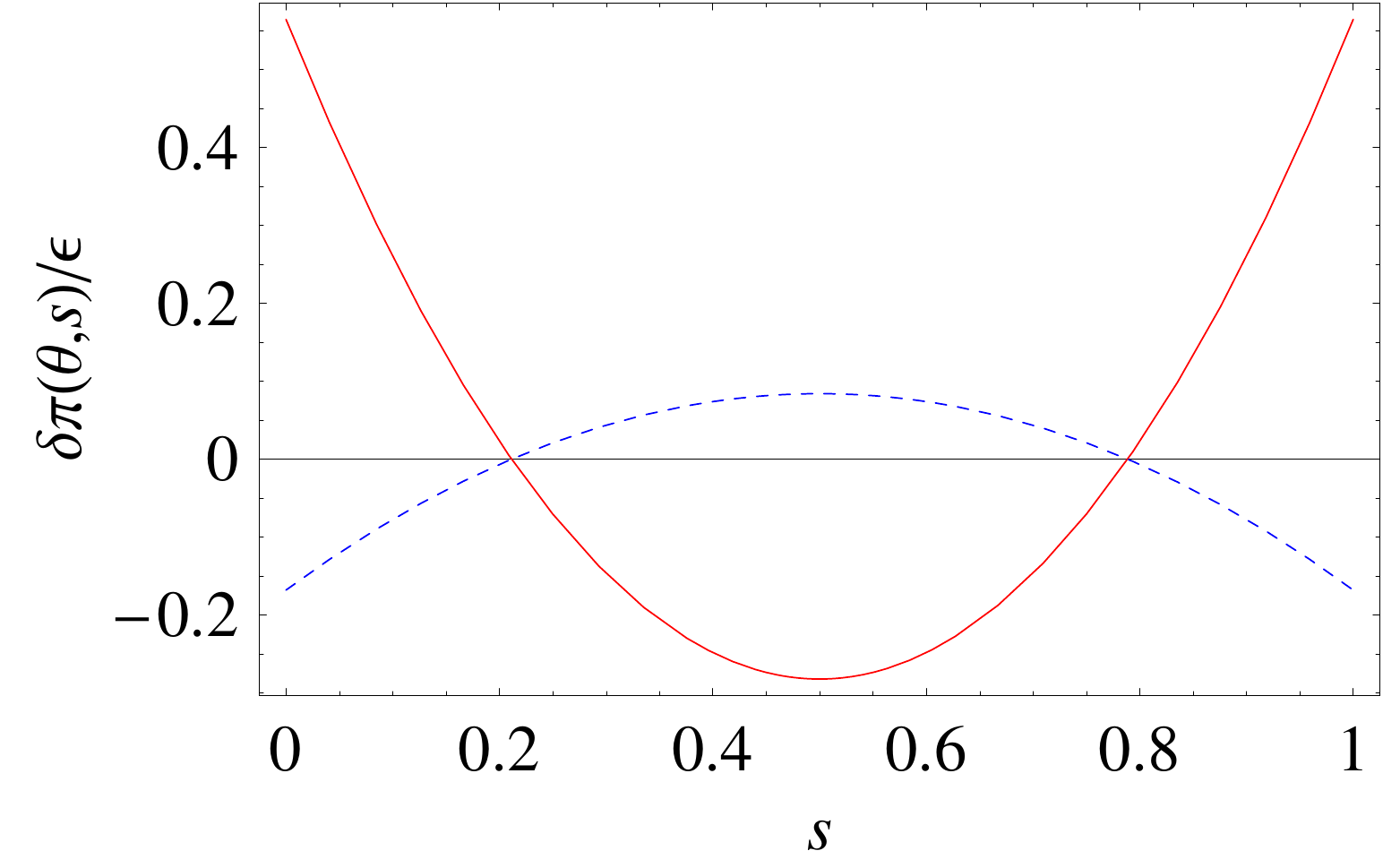}
\caption{(Color online) Height-dependent part, $\delta \pi(\theta,s)$, of   the
  distribution of inverse localization length, $\theta$
  \emph{vs.}~the normalized system height $s=z/L$, for $\theta=1/2$
  (solid, red) and $\theta=2$ (dashed, blue). The cross-link extent is
  $a^2=0$ for both graphs. As in Fig.~\ref{color_distrib}, $\delta\pi(\theta,s)$ is scaled with $\epsilon$.}
\label{fig:distrib(s)}
\end{figure}

To get a better understanding of this anisotropy, we show in
Fig.~\ref{fig:distrib(s)} cuts of Fig.~\ref{color_distrib} for two
fixed $\theta$ values. As one can see, large localization lengths
(such as $\theta = 1/2$, solid red curve) are favored at the boundaries
($s\approx 0,1$), and small localization lengths (such as $\theta = 2$,
dashed blue curve) in the middle of the sample ($s \approx 1/2$). This
behavior is reasonable, since the ends of the chains are more loose. A chain segment close to the top (bottom) boundary has a lower
probability to have a cross-link above (below) and hence is on average
less localized than a chain segment in the middle.


\section{V. Shear Modulus}

In the gel phase, the DPs are localized in the $xy$ plane. Hence the
symmetry with respect to translations in the $x$ and $y$ directions is
spontaneously broken as indicated by a nontrivial expectation value of
the local density as defined in Eq.~(\ref{local_density}). The symmetry breaking occurs
on a local level only, while the macroscopic system (averaged over the
cross-link related disorder) remains homogeneous. We expect low-energy Goldstone
fluctuations and a finite stiffness to static shear deformations in
the $xy$ plane (see Fig.~\ref{schematic}).
In the replica formalism, the overall macroscopic translational
invariance is reflected in common translations of all $n+1$ replicas,
whereas replica-dependent translations generate a family of order
parameters which all give rise to the same free energy.


To investigate the response of the system to shear deformations, we
start from the replica free energy [Eq.~(\ref{free_energy})] and
consider fluctuations around the saddle point (\ref{eq:dp:ansatzQ}),
which correspond to long-wavelength shear deformations ${\bf
  u}^{\alpha}({\bf r})$ in each replica $\alpha=1,..., n$ except the
zeroth replica \cite{SU_EPL,GoldstoneFluct,GoldstoneFluctEPL}. The
latter represents the preparation state before the cross-linking
process, which takes place in a state without shear deformations. We
furthermore want to consider pure shear only and hence require that
volume is conserved, $\nabla\cdot {\bf u}^{\alpha}=0$, and no tilt deformations
are excited; i.e., ${\bf u}^{\alpha}({\bf r})$ are chosen to be
independent of $z$. 

The order parameter for the deformed state thus reads as
\begin{align}
\Omega_u(\hat{q},z) &= (1-Q) \delta_{\hat{q},\hat{0}} +  \nonumber \\
&Q \int\!\! \frac{d^2 r}{A
} \exp\bigl(i {\bf q}_\parallel \cdot {\bf r} + i\sum_{\alpha=1}^n
{\bf q}_\perp^{\alpha}\cdot 
{\bf u}^{\alpha}({\bf r}) \bigr) \nonumber \\
&\times \int_0^\infty \!\! d \xi^2  \mathcal{P}(\xi^2,z) 
\exp\!\left( - \frac{\hat{q}^2 \xi^2}{2}  \right) 
\label{eq:dp:ansatzQ(u)}\;,
\end{align}
where ${\bf q}_\perp^{(\alpha)} := {\bf
  q}^{(\alpha)} - \frac{1}{n+1} {\bf q}_\parallel $.
If the deformations are taken to be spatially uniform we recover the general
solution of the saddle-point equation. Fluctuations around the saddle-point value are incorporated by nonzero $\partial_{x} {\bf u}$ and $\partial_{y} {\bf u}$. These are assumed to be small, corresponding to long-wavelength excitations.


We plug the ansatz (\ref{eq:dp:ansatzQ(u)}) into the free energy
(\ref{free_energy}) and only keep the lowest order in $Q$ and in the
derivatives $\partial_{x} {\bf u}$ and $\partial_{y} {\bf u}$.  Higher
order derivatives, such as $\partial^2_{x,y} {\bf u}$, are
neglected. The result has the form of an elastic free energy of an
incompressible medium,
\begin{align}
  f_{n+1}(\Omega_u) &= f_{\rm sp} + \frac{G}{2N} \int d^2 r
  \sum_{\mu,\nu=1}^{2} \sum_{\alpha=1}^n \left(\frac{\partial u_{\nu}^{\alpha}}{\partial r_\mu} \right)^{\!2} \;.
\end{align}
There $f_{\rm sp}$ is the saddle-point value of the free
energy and $G$ is the shear modulus,
\begin{align}
  G = \left(\frac{Q^2 (\mu^2-1)}{2} - \frac{Q^3}{6} \right)
  \frac{N}{A} k_{\rm B} T \;.
\end{align}
With the distance from the sol-gel transition $\epsilon = \mu^2 - 1$,
the relation $Q=2\epsilon + {\cal O}(\epsilon^2)$ found in the
previous section, and the areal density $n_0 := N/A$ of the polymer
chains, the shear modulus simplifies to
\begin{align}
 G = \frac{2}{3} \epsilon^3 \, n_0 \, k_{\rm B} T \;. 
\end{align}
The scaling of the shear modulus $G \propto \epsilon^3$ close to the
sol-gel transition is in agreement with previous results for isotropic systems
\cite{SU_EPL,GoldstoneFluct,GoldstoneFluctEPL}. This result for the
shear modulus is universal and does not depend on the microscopic
length or energy scales which characterize the polymers and the cross-links.

\section{VI. Conclusions and Outlook}

We have addressed the effect of random permanent cross-links on an array of
directed polymers confined between two planes with their end points free
to slide on them. The cross-links are assumed to have negligible
extent in the preferred direction of the DPs (local in the $z$
direction), but they are spring-like in the transverse ($xy$)
plane. The constraints imposed by the cross-links are treated as
quenched disorder which follows the Deam-Edwards distribution. 

At a certain critical cross-link density, there is a
continuous gelation transition from a sol phase where the DPs are free to
wander in the $x$ and $y$ directions to a gel characterized by the emergence of 
finite localization lengths for the polymer segments which belong to
the infinite percolating cluster. Unlike other isotropic polymer systems which undergo
a similar transition, the DPs are inherently anisotropic and this is
reflected in the height ($z$) dependence of the order parameter and
the associated distribution of localization length. Because of the 
finite extent of the system in the preferred direction, larger localization lengths are favored
closer to the boundaries where the polymer end points are free to slide. 

The gelation transition is accompanied by the emergence of a finite
shear modulus. Our result for the array of cross-linked DPs close to
the gel point agrees with previous results for isotropic systems thus
suggesting universality. As far as in-plane localization and the
relevant shear modulus are concerned,
our system can be viewed as effectively two dimensional. It
is well known that truly long-ranged positional order cannot exist in
two dimensions \cite{Nelson_book}. In \cite{GoldstoneFluct}, it was shown for isotropic
systems that fluctuations drive the order parameter to zero as
expected from the Mermin-Wagner theorem. Yet a quasi-amorphous solid
state survives. It is characterized by a finite stiffness to static
shear deformations and algebraically decaying correlations.

The asymmetry of our system due to the preferred direction of the DPs
entails the existence of a tilt modulus which is different and
independent from the shear modulus. We have only considered
the simplest case of cross-links with negligible extent
in the preferred direction of the DPs.  Using a ``Galilean''
invariance argument, we have shown that cross-links of this type leave the tilt modulus of the
uncross-linked system completely unaffected. We expect
cross-links which connect polymer segments at different heights to
induce an effective ``nonlocal in $z$'' interaction between the
connected polymers. By analogy to a similar interaction in the case of
flux lines in type-II superconductors \cite{tilt}, we can expect the
breaking of the ``Galilean'' invariance to cause an upward renormalization (stiffening) of
the tilt modulus. This putative renormalization may be useful to
quantify the cross-link induced collapse of polymer brushes. We hope to
report on these issues in a future publication.

\section{Acknowledgements}
We gratefully acknowledge financial support by the DFG through Grant No.\ SFB\,602.
P.B.~acknowledges support during the later part of this work by
EPSRC-GB via the University of Cambridge TCM Programme Grant.



\begin{thebibliography}{}

\bibitem{Kardar_book}
M. Kardar, {\it Statistical Physics of Fields}, (Cambridge University Press, Cambridge, 2007).

\bibitem{Nelson_book}
D. R. Nelson, {\it Defects and Geometry in Condensed Matter Physics},
(Cambridge University Press, Cambridge, 2002).

\bibitem{Kamien}
R. D. Kamien, P. Le Doussal, and D. R. Nelson, Phys. Rev. A {\bf 45}, 8727 (1992).

\bibitem{Blatter}
G. Blatter {\it et al.}, Rev. Mod. Phys. {\bf 66}, 1125 (1994).


\bibitem{Hwa}
T. Hwa, Nature {\bf 399}, 17 (1999).


\bibitem{Podgornik}
R. Podgornik, D. C. Rau, and V. A. Parsegian, Macromolecules {\bf 22},
1780 (1989).


\bibitem{Chandra}
S. Chandrasekhar, B. K. Sadashiva, and K. A. Suresh, Pramana {\bf 9},
471  (1977).


\bibitem{Safran}
S. A. Safran, L. A. Turkevich, and P. Pincus, J. Physique Lett. {\bf 45}, 69 (1984).



\bibitem{Rosen} R. E. Rosensweig, {\it Ferrohydrodynamics} (Cambridge University Press, Cambridge, 1985).


\bibitem{Halsey}
T. C. Halsey and W. Toor, Phys. Rev. Lett. {\bf 65}, 2820 (1990).



\bibitem{deGennes_brush}
P. G. de Gennes, Macromolecules {\bf 13}, 1069 (1980).




\bibitem{Halperin_Zhulina}
A. Halperin and E. B. Zhulina, Macromolecules {\bf 24}, 5393 (1991).



\bibitem{Loveless}
D. M. Loveless {\it et al.},
Angew. Chem., Int. Ed. {\bf 45}, 7812 (2006).


\bibitem{Xu}
P. Xu {\it et al.}, 
Biomacromolecules {\bf 5}, 1736 (2004). 


\bibitem{DE}
R. T. Deam and S. F. Edwards, Philos. Trans. R. Soc. London, Ser. A {\bf 280}, 317 (1976).



\bibitem{Castillo}
P. M. Goldbart, H. Castillo, and A. Zippelius, Adv. Phys. {\bf 45},
393 (1996).

\bibitem{SU_EPL}
S. Ulrich, X. Mao, P. M. Goldbart, and A. Zippelius,
Europhys. Lett. {\bf 76}, 677 (2006). 



\bibitem{Pfahl}
X. Xing, S. Pfahl, S. Mukhopadhyay, P. M. Goldbart, and A. Zippelius,
Phys. Rev. E {\bf 77}, 051802 (2008).


\bibitem{p-beine}
P. M. Goldbart and A. Zippelius, Europhys. Lett. {\bf 27}, 599 (1994).


\bibitem{PB_PRL}
P. Benetatos and A. Zippelius, Phys. Rev. Lett. {\bf 99 },  198301   (2007).

\bibitem{Pan_Rab1}
S. V. Panyukov and Y. Rabin, Phys. Rep. {\bf 269}, 1 (1996).

\bibitem{Pan_Rab2}
S. V. Panyukov and Y. Rabin, in {\it Theoretical and Mathematical
  Methods in Polymer Research}, edited by A. Y. Grosberg (Academic Press, New York, 1998).

\bibitem{deGennes}
P. G. de Gennes, in {\it Polymer Liquid Crystals}, edited by A. Ciferri,
W. R. Kringbaum, and R. B. Meyer (Academic Press, New York, 1982), p. 115.


\bibitem{deG-Mat}
P. G. de Gennes and J. Matricon, Rev. Mod. Phys. {\bf 36}, 45 (1964).


\bibitem{TN}
U. C. T\"auber and D. R. Nelson, Phys. Rep. {\bf 289}, 157 (1997).


\bibitem{GoldstoneFluctEPL}
S. Mukhopadhyay, P. M. Goldbart, and A. Zippelius, Europhys. Lett. {\bf 67}, 49 (2004).


\bibitem{GoldstoneFluct}
P. M. Goldbart, S. Mukhopadhyay, and A. Zippelius, Phys. Rev. B {\bf 70}, 184201 (2004).

\bibitem{tilt}
P. Benetatos and M. C. Marchetti, Phys. Rev. B {\bf 59}, 6499 (1999).



\end{thebibliography}
\end{document}